\documentclass[journal,onecolumn]{IEEEtran}

\usepackage{mathrsfs}
\usepackage{color}
\usepackage{enumerate}
\usepackage{mathrsfs}
\usepackage{amssymb}
\usepackage{amsfonts}
\usepackage{amsmath}
\usepackage{multirow}
\usepackage{url}
\usepackage{longtable}
\usepackage{array}
\newtheorem{theorem}{Theorem}[section]
\newtheorem{definition}{Definition}[section]
\newtheorem{lemma}[theorem]{Lemma}

\newtheorem{remark}{Remark}[section]
\newtheorem{Hypothesis}{Hypothesis}[section]

\begin{document}

\title{On the Existence of Certain Optimal Self-Dual Codes with Lengths Between $74$ and $116$}

\author{Tao Zhang, Jerod Michel, Tao Feng and Gennian Ge
\thanks{The research of T. Feng was supported by Fundamental Research Fund for the Central Universities of China,
Zhejiang Provincial Natural Science Foundation under Grant LQ12A01019, the National Natural Science Foundation
of China under Grant 11201418, and the Research Fund for Doctoral Programs from the Ministry of Education of
China under Grant 20120101120089. The research of G. Ge was supported by the National Natural Science Foundation of China under Grant No.~61171198 and Zhejiang Provincial Natural Science Foundation of China under Grant No.~LZ13A010001.}
\thanks{T. Zhang is with the Department of Mathematics, Zhejiang University,
Hangzhou 310027,  China (e-mail: tzh@zju.edu.cn).}
\thanks{J. Michel is with the Department of Mathematics, Zhejiang University,
Hangzhou 310027,  China (e-mail: samarkand\_city@126.com).}
\thanks{T. Feng is with the Department of Mathematics, Zhejiang University,
Hangzhou 310027,  China (e-mail: tfeng@zju.edu.cn). He is also with Beijing Center for Mathematics and Information Interdisciplinary Sciences, Beijing, 100048, China.}
\thanks{G. Ge is with  the School of Mathematical Sciences, Capital Normal University,
Beijing 100048, China (e-mail: gnge@zju.edu.cn). He is also with Beijing Center for Mathematics and Information Interdisciplinary Sciences, Beijing, 100048, China.}
}

\maketitle

\begin{abstract}
The existence of optimal binary self-dual codes is a long-standing research problem. In this paper, we present some results concerning the decomposition of binary self-dual codes with a dihedral automorphism group $D_{2p}$, where $p$ is a prime. These results are applied to construct new self-dual codes with length $78$ or $116$. We obtain $16$ inequivalent self-dual $[78,39,14]$ codes, four of which have new weight enumerators. We also show that there are at least $141$ inequivalent self-dual $[116,58,18]$ codes, most of which are new up to equivalence. Meanwhile,  we give some restrictions on the weight enumerators of singly even self-dual codes. We use these restrictions to exclude some possible weight enumerators of self-dual codes with lengths $74$, $76$, $82$, $98$ and $100$.
\end{abstract}


\section{Introduction}
Binary self-dual codes have been of particular interest for some time now. The extended Hamming $[8,4,4]$ code, the extended Golay $[24,12,8]$ code and certain extended quadratic residue codes are well-known examples of binary self-dual codes. It is known \cite{HP03} that if there is a natural number $r>1$ that divides the weight of all vectors in a binary self-dual code $C$, then $r=2$ or $4$. A binary self-dual code in which all weights are divisible by $4$ is called a doubly even self-dual (or Type II) code, otherwise we call it a singly even self-dual (or Type I) code. All doubly even self-dual codes of length up to $40$ have been classified \cite{P72}, \cite{PS75}, \cite{CPS92}, \cite{BHM12} and a classification of singly even self-dual codes of length up to $38$ is also known \cite{P72}, \cite{PS75}, \cite{CPS92}, \cite{BV02}, \cite{B06}, \cite{HM12}, \cite{BB12}.

 Let $C$ be a binary self-dual code of length $n$ and minimum distance $d$. By results of Mallows-Sloane \cite{MS73} and Rains \cite{R98}, we have
  \[d\leq\begin{cases}4\lfloor\frac{n}{24}\rfloor+4;\textup{ if } n\not\equiv22\pmod{24},\\
4\lfloor\frac{n}{24}\rfloor+6;\textup{ if }n\equiv22\pmod{24}.\end{cases}\]
The code $C$ is called extremal if the above equality holds. If $d=4\lfloor\frac{n}{24}\rfloor+2$ and $n\not\equiv22\pmod{24}$ or if $d=4\lfloor\frac{n}{24}\rfloor+4$ and $n\equiv22\pmod{24}$ then we say $C$ is near extremal. If there is no extremal code with a given length, then we are interested in the code that attains the largest possible minimum distance. Such a code is called an optimal code. A list of possible weight enumerators of extremal self-dual codes of length up to $72$ was given by Conway and Sloane in \cite{CS90}. This list was extended by Dougherty, Gulliver, and Harada in \cite{DGH97}, where lengths are listed up to $100$. However, the existence of some extremal self-dual codes is still unknown. For the classification and enumeration of binary self-dual codes, a survey of known results can be found in \cite{H05}, \cite{RS98}. For the database of self-dual codes, we refer the reader to \cite{HM}, \cite{GO}.

For self-dual codes with large length, a complete classification seems to be impossible. Researchers have focused on self-dual codes with the largest possible minimum weights. Many methods have been proposed to find new self-dual codes with good parameters. Searching for such codes with a double circulant form is a very efficient way, which has led to many good codes \cite{GH98}, \cite{GH06}, \cite{HGK98}. Harada \cite{H97} developed a method involving the double extension of codes. Gaborit and Otmani \cite{GO03} gave a general experimental method to construct self-dual codes. Huffman \cite{H82} constructed binary self-dual codes by applying the automorphism of codes.

 In recent years there have been extensive efforts on the construction of self-dual codes by prescribing certain automorphisms. In $1982$, Huffman \cite{H82} investigated binary self-dual codes with automorphisms of odd prime order and derived the decomposition of such a code as a direct sum of two subcodes. In $1983$, Yorgov \cite{Y83} improved this method and derived necessary and sufficient conditions for codes to be equivalent. In $1997$, Buyuklieva \cite{B1997} developed a new method for constructing binary self-dual codes having an automorphism of order $2$. In $2004$, Dontcheva et al. \cite{DZD04} extended the results to the decomposition of binary self-dual codes possessing an automorphism of order $pq$, where $p$ and $q$ are odd prime numbers. This technique yields many extremal or optimal codes which possess an automorphism (see \cite{BYK12}, \cite{BYR07}, \cite{BYR11}, \cite{DH2003}, \cite{Y14}, \cite{YL13}, \cite{YL2013}, \cite{Y98}).

Let $C$ be a singly even self-dual $[n,n/2,d]$ code and let $C_{0}$ be its doubly even subcode that contains all the codewords of weight divisible by $4$. There are three cosets $C_{1}$, $C_{2}$, $C_{3}$ of $C_{0}$ such that $C_{0}^{\bot}=C_{0}\bigcup C_{1}\bigcup C_{2}\bigcup C_{3}$ and $C=C_{0}\bigcup C_{2}$. The set $S=C_{1}\bigcup C_{3}$ is called the shadow of $C$. Concerning the weight enumerator for $S$, the following theorem was given in \cite{CS90}.
\begin{theorem}\label{shadow}
 Let $S(y)=\Sigma_{r=0}^{n}B_{r}y^{r}$ be the weight enumerator of $S$. Then the following hold:
\begin{enumerate}
\item $B_{r}=B_{n-r}$ for all $r$,
\item $B_{r}=0$ unless $r \equiv n/2\pmod{4}$,
\item $B_{0} = 0$,
\item $B_{r} \leq 1$ for $r < 2n/d$,
\item at most one of $B_{r}$ is nonzero for $r<(d+4)/2$.
\end{enumerate}
\end{theorem}

 It was shown in \cite{DGH97}, \cite{HM06}, \cite{GH06} that the weight enumerator of a binary self-dual $[78,39,14]$ code and its shadow weight enumerator have one of the forms
$$W_{78,1}=1+(3705+8\beta)y^{14}+(62244+512\alpha-24\beta)y^{16}+(774592-4608\alpha-64\beta)y^{18}+\cdots,$$
$$S_{78,1}=\alpha y^{7}+(-\beta-16\alpha)y^{11}+(14\beta+120\alpha+31616)y^{15}+(-560-91\beta+4892160)y^{19}+\cdots,$$
with $\alpha=0,1,2$ and $-448\leq\beta\leq0$, or
$$W_{78,2}=1+(3705+8\alpha)y^{14}+(71460-24\alpha)y^{16}+(658880-64\alpha)y^{18}+\cdots,$$
$$S_{78,2}=y^{3}+(-\alpha-135)y^{11}+(32960+14\alpha)y^{15}+(4885140-91\alpha)y^{19}+\cdots,$$
with $-468\leq\alpha\leq-135$.

Known results on the binary self-dual $[78,39,14]$ codes are listed as follows.
\begin{itemize}
  \item The existence of such codes with the weight enumerator of the form $W_{78,1}$ with $\alpha=0$ and $\beta=-19$ was asserted in \cite{BY03}.
  \item It was shown in \cite{GH06} that there are exactly six inequivalent double circulant self-dual $[78,39,14]$ codes. Five of them have weight enumerators of the form $W_{78,1}$ with $\alpha=0$ and $\beta=0$. The remaining one has weight enumerator of form $W_{78,1}$ with $\alpha=0$ and $\beta=-78$.
  \item Gaborit and Otmani \cite{GO03} constructed a code having weight enumerator of form $W_{78,1}$ with $\alpha=0$ and $\beta=-26$.
  \item In \cite{GHK03}, Gulliver, Harada, and Kim constructed more than $50$ inequivalent codes. Among these codes, one has weight enumerator of form $W_{78,1}$ with $\alpha=0$ and $\beta=-78$, one has weight enumerator of form $W_{78,2}$ with $\alpha=-135$, and all the others have weight enumerators of form $W_{78,1}$ with $\alpha=0$ and $\beta=0$.
\end{itemize}

We also summarize known results on binary self-dual $[116,58,18]$ codes.
\begin{itemize}
  \item Gaborit and Otmani \cite{GO03} constructed a self-dual $[116,58,18]$ code.
  \item Yorgova and Wassermann \cite{YW08} found that there are at least $7$ inequivalent self-dual $[116,58,18]$ codes with an automorphism of order $23$.
\end{itemize}

In this paper, we investigate binary self-dual codes with a dihedral automorphism group $D_{2p}$ of order $2p$, where $p$ is an odd prime. The results will be applied to classify all binary self-dual $[78,39,14]$ codes with a dihedral automorphism group $D_{38}$. Some of these have weight enumerator of form $W_{78,1}$ with $\alpha=0$ and $\beta=-38$ (the existence of such codes was previously unknown). Furthermore, we will show that there exist at least $141$ inequivalent binary self-dual $[116,58,18]$ codes with dihedral automorphism group $D_{58}$. Since the order of the automorphism group is $58$ for all of these codes, almost all of them are new up to equivalence.

In \cite{BW12}, Bouyuklieva and Willems introduced the definition of singly even self-dual codes with minimal shadow.
\begin{definition}We say a self-dual code $C$ of length $n=24m+8l+2r$ with $l=0,1,2$, $r=0,1,2,3$, is a code with minimal shadow if:
\begin{enumerate}
  \item $wt(S)=r$ if $r>0$; and
  \item $wt(S)=4$ if $r=0$.
\end{enumerate}
\end{definition}
They proved that extremal self-dual codes of lengths $n=24m+2,\ 24m+4,\ 24m+6,\ 24m+10,$ and $24m+22$ with minimal shadow do not exist. Moreover, they give explicit bounds in case the shadow is minimal. In this work, we consider extremal self-dual codes with near minimal and near near minimal shadow, and near extremal self-dual codes with minimal, near minimal, and near near minimal shadow and show nonexistence of such codes for certain parameters.

This paper is organized as follows. In Section~\ref{pre} we first recall some results about binary self-dual codes having an automorphism of odd prime order. Then we extend these results to the case where the codes have dihedral automorphism group $D_{2p}$. In Section~\ref{newcode} we investigate self-dual $[78,39,14]$ codes with dihedral automorphism group $D_{38}$ and $[116,58,18]$ codes with dihedral automorphism group $D_{58}$. In Section~\ref{non} we prove nonexistence of self-dual codes for certain parameters. Section~\ref{cl} concludes the paper.

\section{Preliminaries}\label{pre}

Let $C$ be a binary code with an automorphism $\sigma$ of odd prime order $p$. If $\sigma$ has $c$ cycles of length $p$ and $f$ fixed points, we say that $\sigma$ is of type $p-(c;f)$. Without loss of generality we may write
$$\sigma=\Omega_{1}\cdots\Omega_{c}\Omega_{c+1}\cdots\Omega_{c+f},$$
where $\Omega_{i}$ is a $p$-cycle for $i=1,2,\cdots,c$, whereas for $i=c+1,\cdots,c+f$, $\Omega_{i}$ is a fixed point. Let $F_{\sigma}(C)=\{v\in C|v\sigma=v\}$ and $E_{\sigma}(C)=\{v\in C| \textup{wt}(v|_{\Omega_{i}})\equiv0\pmod{2},i=0,1,\cdots,c+f\},$ where $v|_{\Omega_{i}}$ is the restriction of $v$ to $\Omega_{i}$. With this notation, we have the following lemma.

\begin{lemma}\label{1}\cite{H82}
$C=F_{\sigma}(C)\oplus E_{\sigma}(C).$
\end{lemma}

Clearly $v\in F_{\sigma}(C)$ if and only if $v\in C$ and $v$ is constant on each cycle. Let $\pi : F_{\sigma}(C)\rightarrow \mathbb{F}_{2}^{c+f}$ denote the map defined by $\pi(v|_{\Omega_{i}})=v_{j}$ for some $j\in\Omega_{i}$ and $i=1,2,\cdots,c+f$. Then $\pi(F_{\sigma}(C))$ is a binary self-dual code \cite{H82}.

By deleting the last $f$ coordinates of $E_{\sigma}(C)$, we obtain a new code, which is denoted by $E_{\sigma}(C)^{*}$. For $v\in E_{\sigma}(C)^{*}$ we identify $v|_{\Omega_{i}}=(v_{0},v_{1},\cdots,v_{p-1})$ with the polynomial $ v_{0}+v_{1}x+\cdots+v_{p-1}x^{p-1}$ from $P$, where $P$ is the set of even weight polynomials in $\mathbb{F}_{2}[x]/(x^{p}-1)$. Thus we obtain the map $\varphi:E_{\sigma}(C)^{*}\rightarrow P^{c}$, where $P^{c}$ denotes the module of all $c$-tuples over $P$. Clearly, $\varphi(E_{\sigma}(C)^{*})$ is a submodule of the $P$-module $P^{c}$. If the multiplicative order of $2$ modulo $p$ is $p-1$, then the polynomial $1+x+x^{2}+\cdots+x^{p-1}$ of $P$ is irreducible over $\mathbb{F}_{2}$. Hence $P$ is an extension field of $\mathbb{F}_{2}$ with identity $e(x)=x+\cdots+x^{p-1}$ and the following result holds.

\begin{lemma}\label{3}\cite{Y83}
Assume that the multiplicative order of $2$ modulo $p$ is $p-1$. Then a binary code $C$ with an automorphism $\sigma$ of odd prime order $p$ is self-dual if and only if the following two conditions hold.
\begin{enumerate}
\item[(a)] $\pi(F_{\sigma}(C))$ is a binary self-dual code of length $c+f$;
\item[(b)]  $\varphi(E_{\sigma}(C)^{*})$ is a self-dual code of length $c$ over the field $P$ under the inner product $u\cdot v=\sum_{i=1}^{c}u_{i}v_{i}^{q}$ for $q=2^{\frac{p-1}{2}}$.
\end{enumerate}
\end{lemma}

To classify the codes, we need additional conditions for equivalence and we use the following lemma.

\begin{lemma}\label{4}\cite{Y87}
The following transformations applied to $C$ lead to equivalent codes with automorphism $\sigma$:
\begin{enumerate}
\item[(a)] a substitution $x\rightarrow x^{t}$ in $\varphi(E_{\sigma}(C)^{*})$ where $t$ is an integer, $1\leq t\leq p-1$;
\item[(b)]  a multiplication of any coordinate of  $\varphi(E_{\sigma}(C)^{*})$ by $x^{t_{j}}$ where $t_{j}$ is an integer, $0\leq t_{j}\leq p-1$, $j=1,2,\cdots,c$;
\item[(c)] a permutation of the first $c$ cycles of $\sigma$;
\item[(d)]a permutation of the last $f$ coordinates of $C$.
\end{enumerate}
\end{lemma}

The next definition gives an invariant of a code which was introduced by Dontcheva and Harada \cite{DH2003}.

\begin{definition}
Let $C$ be a binary self-dual $[n,k,d]$ code and $\{c_{1},c_{2},\cdots,c_{m}\}$ be the set of all codewords of weight $d$. The intersection numbers of the code $C$ are defined as
$$I_{j}=\sharp\{(c_{s},c_{t})|\textup{dis}(c_{s},c_{t})=j, 1\leq s<t\leq m\},$$
where $\textup{dis}(c_{s},c_{t})$ denotes the Hamming distance between $c_{s}$ and $c_{t}$. Then $I_{j}$ is an invariant under permutations of the coordinates.
\end{definition}

The following two lemmas are efficient in excluding some types of automorphisms of a self-dual code.

\begin{lemma}\label{6}\cite{Y87}
Let $C$ be a binary self-dual $[n,k,d]$ code and let $\sigma\in Aut(C)$ be of type $p-(c;f)$, where $p$ is an odd prime. If $g(s)=\sum_{i=0}^{s-1}\lceil\frac{d}{2^{i}}\rceil$, then
\begin{enumerate}
\item[(a)] $pc\geq g(\frac{p-1}{2}c)$, and
\item[(b)] $f\geq g(\frac{f-c}{2})$ for $f>c$.
\end{enumerate}
\end{lemma}

\begin{lemma}\label{7}\cite{BMW10}
Let $C$ be a binary self-dual code of length $n$ and let $\sigma$ be an automorphism of $C$ of type $p-(c;f)$, where $p$ is an odd prime. If the multiplicative order of $2$ modulo $p$ is even, then $c$ is even.
\end{lemma}

In order to get our results, we give the following hypothesis.
\begin{Hypothesis}\label{hy}
$C$ is a binary self-dual $[n,n/2,d]$ code, where $n\geq52$, $n=4p+f$, $p$ is an odd prime number with $2$ as a primitive root, $f=0,2,4$ and
\[d\geq\begin{cases}4\lfloor\frac{n}{24}\rfloor+2;\textup{ if } n\not\equiv22\pmod{24},\\
4\lfloor\frac{n}{24}\rfloor+4;\textup{ if }n\equiv22\pmod{24},\end{cases}\]
then $p\geq13$ and $d\geq10$.
\end{Hypothesis}

As a preparation, we have the following lemma.

\begin{lemma}\label{423}
Under Hypothesis~\ref{hy}, if $C$ has an automorphism $\sigma$ of type $p-(4;f)$, then $\varphi(E_{\sigma}(C)^{*})$ is a self-dual $[4,2,3]$ code over the field $P\cong\mathbb{F}_{2^{p-1}}$.
\end{lemma}
\begin{proof}
According to Lemma ~\ref{3}, $\varphi(E_{\sigma}(C)^{*})$ is a self-dual $[4,2]$ code over the field $P\cong\mathbb{F}_{2^{p-1}}$. Since the minimum distance of $\varphi(E_{\sigma}(C)^{*})$ cannot be $4$, we only need to prove that $\varphi(E_{\sigma}(C)^{*})$ has minimum distance $\neq1,2$.

 \textbf{Case 1: $\varphi(E_{\sigma}(C)^{*})$ has minimum weight $1$.}

 Take $\textbf{u}\in\varphi(E_{\sigma}(C)^{*})$ with $\textup{wt}(\textbf{u})=1$. Then we can assume that $\textbf{u}=(v_{1},0,0,0)$ with $v_{1}\neq0$. Since $(x+1)v_{1}^{-1}\textbf{u}=(x+1,0,0,0)\in\varphi(E_{\sigma}(C)^{*})$, we have $\textup{wt}(\varphi^{-1}((x+1)v_{1}^{-1}\textbf{u}))=2$ which contradicts the fact $d\geq10$.

 \textbf{Case 2: $\varphi(E_{\sigma}(C)^{*})$ has minimum weight $2$.}

 Take $\textbf{u}\in\varphi(E_{\sigma}(C)^{*})$ with $\textup{wt}(\textbf{u})=2$. Suppose $\textbf{u}=(v_{1},v_{2},0,0)$ with $v_{1},v_{2}\neq0$. Let $U=\{v\textbf{u}|v\in P\}$. Then $\textup{dim}_{\mathbb{F}_{2}}U=p-1$. Set $W=\varphi^{-1}(U)\subseteq E_{\sigma}(C)^{*}$. Let $W^{*}$ be the code obtained from $W$ by deleting the last $2p$ coordinates. Then $W^{*}$ is a $[2p,p-1,d]$ code. To get a contradiction, take $g(s)=\sum_{i=0}^{s-1}\lceil\frac{d}{2^{i}}\rceil.$

 First we consider the case $p\equiv1\pmod{6}$ and $f=0$. We can write $p=6k+1$, for some integer $k\geq2$. Then $n=24k+4$, $d\geq4k+2$, $g(1)\geq4k+2$, $g(2)\geq6k+3$ and $g(3)\geq7k+4$. If $2^{l}<2k+1\leq2^{l+1}$ for $l\in\mathbb{N}$ then for $i>l$ we have $\frac{2k+1}{2^{i}}\leq2^{l+1-i}\leq1$ and therefore $\lceil\frac{2k+1}{2^{i}}\rceil=1$. Hence
 \begin{align*}
g(p-1) &\geq \sum_{i=0}^{p-2}\lceil\frac{4k+2}{2^{i}}\rceil\geq 7k+4+\sum_{i=2}^{p-3}\lceil\frac{2k+1}{2^{i}}\rceil \\
          &= 7k+4+\sum_{i=2}^{l}\lceil\frac{2k+1}{2^{i}}\rceil+(p-3-l)\\
          &= \sum_{i=2}^{l}\lceil\frac{2k+1}{2^{i}}\rceil+12k+2+(k-l).\\
\end{align*}

If $k=2$, then $l=2$ and $g(p-1)>12k+2.$

If $k=3$, then $l=2$ and $g(p-1)>12k+2.$

If $k\geq4$, then $l\geq3$. Since $(k-l)>(2^{l-1}-l-\frac{1}{2})>0$, we get $g(p-1)>12k+2.$ Consequently, $g(p-1)>12k+2=2p$ which contradicts the Griesmer Bound \cite{HP03}.

For the other cases of $p$ and $f$, a similar discussion leads to a contradiction. Hence, $\varphi(E_{\sigma}(C)^{*})$ can not have minimum weight $2$.
\end{proof}

Now we are ready to prove our result.
 \begin{theorem}\label{8}
Under Hypothesis~\ref{hy}, if $C$ has a dihedral automorphism group $D_{2p}$, and $\sigma\in D_{2p}$ is an automorphism of type $p-(4;f)$ then $C=F_{\sigma}(C)\oplus E_{\sigma}(C)$, and there is a generator matrix of $\varphi(E_{\sigma}(C)^{*})$ that has the form
\begin{equation}
\textup{gen}(\varphi(E_{\sigma}(C)^{*}))=\left[ \begin {array}{cccc} {b^{u_{1}}} & {0}&{a^{v_{1}}}&{a^{v_{2}}b^{u_{3}}}\\ \noalign{\medskip}{0}&{b^{u_{2}}}&{a^{v_{2}}}&{a^{v_{1}}b^{u_{3}}}\end {array} \right],
\end{equation}
where $a,\ b$ are the elements of $P$ of order $q-1$ and $\frac{q+1}{p}$, respectively. And $a^{v_{1}}+a^{v_{2}}=e$, $1\leq v_{1}<v_{2}\leq q-2$, $0\leq u_{i}\leq \frac{q+1}{p}-1$ for $i=1,2,3$, where $q=2^{\frac{p-1}{2}}$. Also, the $u_{i}$'s satisfy one of the following conditions:
\begin{enumerate}
  \item $u_{1}+u_{2}\equiv u_{3} \pmod{\frac{q+1}{p}};$
  \item $u_{2}+u_{3}\equiv u_{1}\pmod{\frac{q+1}{p}};$
  \item $u_{1}+u_{3}\equiv u_{2}\pmod{\frac{q+1}{p}};$
  \item $u_{1}=u_{2}=u_{3}=0.$
\end{enumerate}
 \end{theorem}
\begin{proof}
Suppose that $C$ is a self-dual $[n,n/2,d]$ code with dihedral automorphism group $D_{2p}$. Let $\sigma\in D_{2p}$ be an automorphism of type $p-(4;f)$. Without loss of generality, we can write
$$\sigma=(1,\cdots,p)(p+1,\cdots,2p)(2p+1,\cdots,3p)(3p+1,\cdots,4p).$$
 Then $\varphi(E_{\sigma}(C)^{*})$ is a self-dual $[4,2,3]$ code over the field $P$ under the inner product $u\cdot v=\sum_{i=1}^{c}u_{i}v_{i}^{q}$ for $q=2^{\frac{p-1}{2}}$ by Lemma \ref{423}. Let $e$ be the identity element of $P$, $\alpha$ a primitive element of $P$, and set $a=\alpha^{q+1}$ and $b=\alpha^{(q-1)p}$. Then by a computation similar to that in \cite{Y87}, we have
\begin{equation}\label{9}
\textup{gen}(\varphi(E_{\sigma}(C)^{*}))=\left[ \begin {array}{cccc} {b^{u_{1}}} & {0}&{a^{v_{1}}}&{a^{v_{2}}b^{u_{3}}}\\ \noalign{\medskip}{0}&{b^{u_{2}}}&{a^{v_{2}}}&{a^{v_{1}}b^{u_{3}}}\end {array} \right],
\end{equation}
where $a^{v_{1}}+a^{v_{2}}=e$, $1\leq v_{1}<v_{2}\leq q-2$, and $0\leq u_{i}\leq \frac{q+1}{p}-1$ for $i=1,2,3.$

We consider the involution of $D_{2p}$ acting on $C$. Let $\tau\in D_{2p}$ be an element of order $2$ such that $\tau\sigma\tau=\sigma^{-1}$, that is
 \begin{equation}
 \begin{split}
&(\tau(1),\cdots,\tau(p))(\tau(p+1),\cdots,\tau(2p))(\tau(2p+1),\cdots,\tau(3p))(\tau(3p+1),\cdots,\tau(4p))\\
&=(p,\cdots,1)(2p,\cdots,p+1)(3p,\cdots,2p+1)(4p,\cdots,3p+1).
\end{split}
\end{equation}
 Then by Lemma~\ref{4}(b)(d) we may relabel the coordinates so that $\tau\in S$ where $S$ is the set consisting of the following elements:
$$(1,2p)\cdots(p,p+1)(2p+1,4p)\cdots(3p,3p+1),$$
 $$(1,3p)\cdots(p,2p+1)(p+1,4p)\cdots(2p,3p+1),$$
 $$(1,4p)\cdots(p,3p+1)(p+1,3p)\cdots(2p,2p+1), $$
 $$(1,p)\cdots(\frac{p-1}{2},\frac{p+3}{2})(p+1,2p)\cdots(\frac{3p-1}{2},\frac{3p+3}{2})\cdots(3p+1,4p)\cdots(\frac{7p-1}{2},\frac{7p+3}{2}),$$
 $$(1,p)\cdots(\frac{p-1}{2},\frac{p+3}{2})(p+1,2p)\cdots(\frac{3p-1}{2},\frac{3p+3}{2})(2p+1,4p)\cdots(3p,3p+1),$$
 $$(1,p)\cdots(\frac{p-1}{2},\frac{p+3}{2})(p+1,3p)\cdots(2p,2p+1)(3p+1,4p)\cdots(\frac{7p-1}{2},\frac{7p+3}{2}),$$
 $$(1,p)\cdots(\frac{p-1}{2},\frac{p+3}{2})(p+1,4p)\cdots(2p,3p+1)(2p+1,3p)\cdots(\frac{5p-1}{2},\frac{5p+1}{2}),$$
 $$(1,2p)\cdots(p,p+1)(2p+1,3p)\cdots(\frac{5p-1}{2},\frac{5p+1}{2})(3p+1,4p)\cdots(\frac{7p-1}{2},\frac{7p+3}{2}),$$
 $$(1,3p)\cdots(p,2p+1)(p+1,2p)\cdots(\frac{3p-1}{2},\frac{3p+3}{2})(3p+1,4p)\cdots(\frac{7p-1}{2},\frac{7p+3}{2}),$$
 $$(1,4p)\cdots(p,3p+1)(p+1,2p)\cdots(\frac{3p-1}{2},\frac{3p+3}{2})(2p+1,3p)\cdots(\frac{5p-1}{2},\frac{5p+1}{2}).$$

 We now consider the action of $\tau$ on $\varphi(E_{\sigma}(C)^{*})$.

 Let $-:\mathbb{F}_{2^{p-1}}\rightarrow\mathbb{F}_{2^{p-1}}, x\rightarrow \overline{x}=x^{q}$ be the nontrivial Galois automorphism of $\mathbb{F}_{2^{p-1}}$ with fixed field $\mathbb{F}_{q}$.

Since the computation of each case is similar, we take $\tau=(1,2p)\cdots(p,p+1)(2p+1,4p)\cdots(3p,3p+1)$ as a sample. For the other cases, we just list the results.

 The action of $\tau$ is given by
 $$\tau(x_{1},x_{2},x_{3},x_{4})=(\overline{x_{2}},\overline{x_{1}},\overline{x_{4}},\overline{x_{3}}),$$
 where $x_{1},x_{2},x_{3},x_{4}\in\mathbb{F}_{2^{p-1}}.$ So
 \begin{equation}\label{5}
\tau(\textup{gen}(\varphi(E_{\sigma}(C)^{*})))=\left[ \begin {array}{cccc} {0} & {\overline{b}^{u_{1}}}&{a^{v_{2}}\overline{b}^{u_{3}}}&{a^{v_{1}}}\\ \noalign{\medskip}{\overline{b}^{u_{2}}}&{0}&{a^{v_{1}}\overline{b}^{u_{3}}}&{a^{v_{2}}}\end {array} \right].
\end{equation}

 Since $\tau\in Aut(C)$, then $\sigma^{-1}(\tau(E_{\sigma}(C)))\subseteq C$, due to the orthogonality of the rows of matrices (\ref{9}) and (\ref{5}), we get the following equations
 $$a^{v_{1}}+a^{v_{2}}=e,\ b^{u_{1}+u_{2}}+b^{u_{3}}a^{2v_{1}}+b^{u_{3}}a^{2v_{2}}=0,$$
 which imply that $u_{1}+u_{2}\equiv u_{3} \pmod{\frac{q+1}{p}}.$

If $\tau=(1,3p)\cdots(p,2p+1)(p+1,4p)\cdots(2p,3p+1)$, then $u_{2}+u_{3}\equiv u_{1}\pmod{\frac{q+1}{p}}.$

If $\tau=(1,4p)\cdots(p,3p+1)(p+1,3p)\cdots(2p,2p+1)$, then $u_{1}+u_{3}\equiv u_{2}\pmod{\frac{q+1}{p}}.$

If $\tau=(1,p)\cdots(\frac{p-1}{2},\frac{p+3}{2})(p+1,2p)\cdots(\frac{3p-1}{2},\frac{3p+3}{2})(2p+1,3p)\cdots(\frac{5p-1}{2},\frac{5p+1}{2})(3p+1,4p)\cdots(\frac{7p-1}{2},\frac{7p+3}{2}))$,
then $u_{1}=u_{2}=u_{3}=0.$

In the other cases, there is no solution.

\end{proof}

\begin{remark}
 Our assumptions may seem restrictive, but they make for simple notations and are sufficient for our purposes.
\end{remark}

\section{New Optimal Self-Dual Codes with Dihedral Automorphism Group $D_{2p}$}\label{newcode}
\subsection{Self-Dual $[78,39,14]$ Codes with Dihedral Automorphism Group $D_{38}$}\label{s78}
\begin{theorem}
There are exactly $16$ inequivalent self-dual $[78,39,14]$ codes with dihedral automorphism group $D_{38}$; they are listed in Table \ref{78}.
\end{theorem}
\begin{proof}
Assume that $C$ is a self-dual $[78,39,14]$ code having dihedral automorphism group $D_{38}$ and let $\sigma\in D_{38}$ be an automorphism of order $19$. It is easy to see that $19-(4;2)$ is the only possible type for $\sigma$ by Lemmas \ref{6} and \ref{7}. By Lemma~\ref{3}, $\pi(F_{\sigma}(C))$ is a binary self-dual $[6,3]$ code. Consequently,
\begin{equation}\label{F}
\textup{gen}(\pi(F_{\sigma}(C)))=\left[ \begin {array}{ccccccc} {1} & {0}&{0}&{1}&{}&{0}&{0}\\ \noalign{\medskip}{0}&{1}&{0}&{0}&{}&{1}&{0}\\ \noalign{\medskip}{0}&{0}&{1}&{0}&{}&{0}&{1}\end {array} \right].
\end{equation}

Let $P$ be the vector space of even weight polynomials in $\mathbb{F}_{2}[x]/(x^{19}-1)$, $e$ be the identity of $P$, $a=x+x^{2}+x^{5}+x^{6}+x^{13}+x^{14}+x^{17}+x^{18}$, and $b=x^{4}+x^{7}+x^{8}+x^{9}+x^{10}+x^{11}+x^{12}+x^{15}+x^{16}+x^{17}$. It is easy to verify that the multiplicative orders of $a$ and $b$ are $2^{9}-1$ and $(2^{9}+1)/19$, respectively.

Since $s(19)=18$, it is easy to verify Hypothesis \ref{hy}. By Theorem~\ref{8} there is a generator matrix of $\varphi(E_{\sigma}(C)^{*})$ of the form
\begin{equation}\label{E}
\textup{gen}(\varphi(E_{\sigma}(C)^{*}))=\left[ \begin {array}{cccc} {b^{u_{1}}} & {0}&{a^{v_{1}}}&{a^{v_{2}}b^{u_{3}}}\\ \noalign{\medskip}{0}&{b^{u_{2}}}&{a^{v_{2}}}&{a^{v_{1}}b^{u_{3}}}\end {array} \right],
\end{equation}
where $a^{v_{1}}+a^{v_{2}}=e$, $1\leq v_{1}<v_{2}\leq 510$, $0\leq u_{i}\leq 26$ for $i=1,2,3$, and the $u_{i}$'s satisfy one of the following conditions:
\begin{enumerate}
  \item $u_{1}+u_{2}\equiv u_{3} \pmod{27};$
  \item $u_{2}+u_{3}\equiv u_{1}\pmod{27};$
  \item $u_{1}+u_{3}\equiv u_{2}\pmod{27};$
  \item $u_{1}=u_{2}=u_{3}=0.$
\end{enumerate}

From \cite{DY03}, we have $(v_{1},v_{2})\in V$, where
$V=\{(1,93), (6,13), (7,505), (9,59), (15,37), (19,105), (20,99), (21,87), (25,251),\\ (29,178), (31,193), (34,175), (39,111), (43,246), (45,61), (46,255), (49,119), (63,190), (73,219), (83,138), (91,167),\\ (94,169), (103,108), (106,239), (114,221), (125,187), (155,213), (179,220), (191,242)\}.$

Let $G$ be the automorphism group of the code generated by $\textup{gen}(\pi(F_{\sigma}(C)))$. Let $S$ be the stabilizer of $G$ on the set of fixed points $\{5,6\}$. Suppose $s$ belongs to the symmetric group $S_{4}$. Then we use $C^{s}$ to denote the self-dual code determined by $E_{\sigma}$ and the matrix $\pi^{-1}(s(\textup{gen}(\pi(F_{\sigma}(C)))))$. By \cite[Lemma 4.1]{K12}, if $s_{1}$ and $s_{2}$ are permutations from the group $S_{4}$ and $Ss_{1}=Ss_{2}$, then the codes $C^{s_{1}}$ and $C^{s_{2}}$ are equivalent. So in order to get all inequivalent self-dual $[78,39,14]$ codes with a dihedral automorphism group $D_{38}$, we must check $\pi^{-1}(s(\textup{gen}(\pi(F_{\sigma}(C)))))$, where
$s\in S_{4}/S=\{I, (1,2,3,4), (1,2), (1,3)(2,4), (1,3,4), (1,4,3,2)\}.$

Now we consider the involution $\tau$ of $D_{38}$ acting on $\pi^{-1}(s(\textup{gen}(\pi(F_{\sigma}(C)))))$.

If $\tau=(1,38)\cdots(19,20)(39,76)\cdots(57,58)$, an easy computation shows that $s$ must be $(1,2,3,4)$.

Similarly,
if $\tau=(1,57)\cdots(19,39)(20,76)\cdots(38,58)$, then $s\in\{(1,3,4),(1,2)\}$.

If $\tau=(1,76)\cdots(19,58)(20,57)\cdots(38,39)$, then $s\in\{I,(1,3)(2,4),(1,4,3,2)\}$.

If $\tau=(1,19)\cdots(9,11)(20,38)\cdots(28,30)(39,57)\cdots(47,49)(58,76)\cdots(66,68)$, then $s\in \{I, (1,2,3,4), (1,2), (1,3)(2,4),\\ (1,3,4), (1,4,3,2)\}$.

Therefore, we should analyze the generator matrix
\begin{equation}
\textup{gen}(C)=\left[ \begin {array}{c} {\pi^{-1}(s(\textup{gen}(\pi(F_{\sigma}(C)))))}\\ \noalign{\medskip}{\textup{gen}(E_{\sigma})}\end {array} \right],
\end{equation}
where $\textup{gen}(\pi(F_{\sigma}(C)))$ has been determined in (\ref{F}) and $\textup{gen}(E_{\sigma})$ corresponds to (\ref{E}) with $(v_{1},v_{2})\in V$, $0\leq u_{i}\leq 26$ for $i=1,2,3$, and the $u_{i}$'s $(i=1,2,3)$ and $s$ satisfy one of the following conditions:
\begin{enumerate}
  \item $u_{1}+u_{2}\equiv u_{3} \pmod{27},s=(1,2,3,4);$
  \item $u_{2}+u_{3}\equiv u_{1}\pmod{27}, s\in\{(1,3,4),(1,2)\};$
  \item $u_{1}+u_{3}\equiv u_{2}\pmod{27},s\in\{I,(1,3)(2,4),(1,4,3,2)\};$
  \item $u_{1}=u_{2}=u_{3}=0,s\in \{I, (1,2,3,4), (1,2), (1,3)(2,4), (1,3,4), (1,4,3,2)\}.$
\end{enumerate}

Using MAGMA \cite{BCP97}, we found exactly $16$ inequivalent self-dual $[78,39,14]$ codes with dihedral automorphism group $D_{38}$. Four of them have weight enumerator $W_{78,1}$ with $\alpha=0$ and $\beta=-38$ which was unknown before. The corresponding values of the parameters are given in Table \ref{78}. All the codes have weight enumerators $W_{78,1}$ with $\alpha=0$, so we just list the values of $\beta$. Here $I_{28}$ is the intersection number. $I$ is the identity permutation in the group $S_{4}$ and $\sharp$Aut denotes the order of the automorphism group of the corresponding code.

Since all the intersection numbers of the codes listed in Table \ref{78} are different, they are inequivalent.

\end{proof}
\begin{table}[h]
\begin{center}
\caption{Self-Dual $[78,39,14]$ Codes with Dihedral Automorphism Group $D_{38}$}
\begin{tabular}{|c|c|c|c|c|c|c|c|c|c|}
\hline
Code  &  $u_{1}$  &  $u_{2}$  & $u_{3}$  & $v_{1}$ & $v_{2}$ & $s$ & $\beta$ & $I_{28}$ & $\sharp$Aut\\ \hline
$C_{1}$ & $6$ & $15$ & $21$ & $1$ & $93$ & $(1,2,3,4)$ & $0$ & $646285$ & $38$\\ \hline
$C_{2}$ & $6$ & $12$ & $18$ & $1$ & $93$ & $(1,2,3,4)$ & $0$ & $643910$ & $38$\\ \hline
$C_{3}$ & $10$ & $10$ & $0$ & $215$ & $335$ & $(1,3,4)$ & $0$ & $644537$ & $38$\\ \hline
$C_{4}$ & $10$ & $10$ & $0$ & $215$ & $335$ & $I$ & $0$ & $646266$ & $38$\\ \hline
$C_{5}$ & $10$ & $13$ & $3$ & $29$ & $178$ & $I$ & $0$ & $643815$ & $38$\\ \hline
$C_{6}$ & $10$ & $34$ & $24$ & $29$ & $178$ & $I$ & $0$ & $642428$ & $38$\\ \hline
$C_{7}$ & $29$ & $9$ & $20$ & $35$ & $231$ & $(1,3,4)$ & $0$ & $642010$ & $38$\\ \hline
$C_{8}$ & $22$ & $13$ & $18$ & $49$ & $119$ & $I$ & $0$ & $645107$ & $38$\\ \hline
$C_{9}$ & $25$ & $21$ & $4$ & $83$ & $138$ & $(1,3,4)$ & $0$ & $650313$ & $38$\\ \hline
$C_{10}$ & $24$ & $2$ & $22$ & $83$ & $138$ & $(1,3,4)$ & $0$ & $647254$ & $38$\\ \hline
$C_{11}$ & $20$ & $25$ & $22$ & $83$ & $138$ & $(1,3,4)$ & $0$ & $645278$ & $38$\\ \hline
$C_{12}$ & $17$ & $21$ & $23$ & $83$ & $138$ & $(1,3,4)$ & $0$ & $648546$ & $38$\\ \hline
$C_{13}$ & $26$ & $6$ & $5$ & $9$ & $59$ & $(1,2,3,4)$ & $-38$ & $547523$ & $38$\\ \hline
$C_{14}$ & $21$ & $12$ & $6$ & $19$ & $105$ & $(1,2,3,4)$ & $-38$ & $546573$ & $38$\\ \hline
$C_{15}$ & $21$ & $15$ & $9$ & $19$ & $105$ & $(1,2,3,4)$ & $-38$ & $546649$ & $38$\\ \hline
$C_{16}$ & $15$ & $5$ & $17$ & $29$ & $178$ & $I$ & $-38$ & $544882$ & $38$\\ \hline
\end{tabular}
\label{78}
\end{center}
\end{table}

\begin{remark}
It took about $5$ hours on a $3$ GHz CPU to classify the self-dual $[78, 39, 14]$ codes with a dihedral automorphism group $D_{38}$.
\end{remark}

\subsection{Self-Dual $[116,58,18]$ Codes with a Dihedral Automorphism Group $D_{58}$}\label{s116}
\begin{theorem}
There are at least $141$ inequivalent self-dual $[116,58,18]$ codes with dihedral automorphism group $D_{58}$. They are listed in Tables \ref{116}.
\end{theorem}
\begin{proof}
Suppose $C$ is a self-dual $[116,58,18]$ code with dihedral automorphism group $D_{58}$ and let $\sigma\in D_{58}$ have order $29$. A similar discussion to that in the previous subsection leads to
\begin{equation}
\textup{gen}(C)=\left[ \begin {array}{c} {\pi^{-1}(s(\textup{gen}(\pi(F_{\sigma}(C)))))}\\ \noalign{\medskip}{\textup{gen}(E_{\sigma}(C))}\end {array} \right],
\end{equation}
where
\begin{equation}\label{F1}
\textup{gen}(\pi(F_{\sigma}(C)))=\left[ \begin {array}{cccc} {1} & {1}&{0}&{0}\\ \noalign{\medskip}{0}&{0}&{1}&{1}\end {array} \right],
\end{equation}
$s\in S_{4}/S$, where $S$ is the automorphism group of the code generated by $\textup{gen}(\pi(F_{\sigma}(C)))$, and $\textup{gen}(E_{\sigma}(C))$ corresponds to
\begin{equation}\label{E1}
\textup{gen}(\varphi(E_{\sigma}(C)^{*}))=\left[ \begin {array}{cccc} {b^{u_{1}}} & {0}&{a^{v_{1}}}&{a^{v_{2}}b^{u_{3}}}\\ \noalign{\medskip}{0}&{b^{u_{2}}}&{a^{v_{2}}}&{a^{v_{1}}b^{u_{3}}}\end {array} \right],
\end{equation}
with $a=x+x^{3}+x^{4}+x^{6}+x^{9}+x^{10}+x^{11}+x^{18}+x^{19}+x^{20}+x^{23}+x^{25}+x^{26}+x^{28}\in P$ of multiplicative order $2^{14}-1$, $b=x+x^{2}+x^{3}+x^{4}+x^{6}+x^{7}+x^{10}+x^{12}+x^{13}+x^{14}+x^{17}+x^{19}+x^{20}+x^{21}+x^{22}+x^{28}\in P$ of multiplicative order $(2^{14}+1)/29$ and $P$ being the set of all even weight polynomials in $\mathbb{F}_{2}[x]/(x^{29}-1)$, $a^{v_{1}}+a^{v_{2}}=e$, $1\leq v_{1}<v_{2}\leq 2^{14}-2$ and $0\leq u_{i}\leq 564$ for $i=1,2,3$. The $u_{i}$'s also satisfy one of the following conditions:
\begin{enumerate}
  \item $u_{1}+u_{2}\equiv u_{3} \pmod{565};$
  \item $u_{2}+u_{3}\equiv u_{1}\pmod{565};$
  \item $u_{1}+u_{3}\equiv u_{2}\pmod{565};$
  \item $u_{1}=u_{2}=u_{3}=0.$
\end{enumerate}

Using MAGMA \cite{BCP97}, we found at least $141$ inequivalent self-dual $[116,58,18]$ codes with dihedral automorphism group $D_{58}$. The corresponding values of the parameters are given in Table \ref{116}. Here $A_{18}$ denotes the number of codewords with weight $18$, and $I_{36}$ is the intersection number. $I$ is the identity permutation in the group $S_{4}$ and $\sharp$Aut denotes the order of the automorphism group of the corresponding code.

It is easy to see that all the intersection numbers of the codes listed in Table \ref{116} are different, hence they are inequivalent. Since all the automorphism groups have order $58$, they are inequivalent with the codes constructed in \cite{YW08}.

\end{proof}

\section{Nonexistence of Some Self-Dual Codes}\label{non}
\subsection{Some Restrictions on Weight Enumerators}

In this section, we study the nonexistence of some self-dual codes. According to \cite{CS90}, if $C$ is a singly-even self-dual code of length $n=24m+8l+2r$ with $l=0,1,2$ and $r=0,1,2,3$, the weight enumerator of $C$ and $S$ are given by:
\[ W(y)=\Sigma_{j=0}^{12m+4l+r}a_{j}y^{2j}=\Sigma_{i=0}^{3m+l}c_{i}(1+y^{2})^{12m+4l+r-4i}(y^{2}(1-y^{2})^{2})^{i}, \]
\[S(y)=\Sigma_{j=0}^{6m+2l}b_{j}y^{4j+r}=\Sigma_{i=0}^{3m+l}(-1)^{i}c_{i}2^{12m+4l+r-6i}y^{12m+4l+r-4i}(1-y^{4})^{2i}. \]
We can write the $c_{i}$ as a linear combination of the $a_{i}$ and as a linear combination of the $b_{i}$ \cite{R98}:
\begin{equation}\label{eq1}
 c_{i}=\Sigma_{j=0}^{i}\alpha_{ij}a_{j}=\Sigma_{j=0}^{3m+l-i}\beta_{ij}b_{j}.
\end{equation}

As a preparation, we give the definition of near minimal shadow and near near minimal shadow.
\begin{definition}
We say a self-dual code $C$ of length $n=24m+8l+2r$ with $l=0,1,2$, $r=0,1,2,3$, is a code with near minimal shadow if:
\begin{enumerate}
  \item $wt(S)=r+4$ if $r>0$; and
  \item $wt(S)=8$ if $r=0$.
\end{enumerate}
And a code with near near minimal shadow if:
\begin{enumerate}
  \item $wt(S)=r+8$ if $r>0$; and
  \item $wt(S)=12$ if $r=0$.
\end{enumerate}
\end{definition}

Then we have the following theorem.
\begin{theorem}\label{th2.1} An extremal self-dual code of length $n=24m+8l+2r$ with near minimal shadow does not exist whenever:
\begin{enumerate}
\item $r=1$ and $l=0$,
\item $r=1$, $l=1$ and $\frac{-12m+5}{-4m-2}\binom{5m+1}{m}-\frac{3m}{2m+1}\binom{5m}{m-1}$ is not an integer,
\item $r=2$, $l=0$ and $\frac{2(6m+1)(8m+1)}{16m(2m+1)}\binom{5m}{m-1}-\frac{3m-1}{2m+1}\binom{5m-1}{m-2}$ is not an integer,
\item $r=3$, $l=0$ and $\frac{3(4m+1)(6m+1)}{8m(2m+1)}\binom{5m}{m-1}-\frac{3m-1}{2m+1}\binom{5m-1}{m-2}$ is not an integer.
\end{enumerate}
\end{theorem}
\begin{proof}
Suppose $C$ is an extremal singly even self-dual code with near minimal shadow so that $d=4m+4$ and $wt(S)=r+4$ if $r=1,2,3$ and $wt(S)=8$ if $r=0$. Then we must have $a_{0}=1,a_{1}=\cdots=a_{2m+1}=0$.

 By Theorem~\ref{shadow}, if $r>0$ we have $b_{0}=0$ and $b_{1}=1$ for $m\geq 1$, and if $r=0$ we have $b_{0}=b_{1}=0$ and $b_{2}=1$ for $m\geq 2$. Also, if $r>0$ and $m\geq 1$, then $b_{2}=b_{3}=\cdots=b_{m-2}=0$, and if $r=0$ and $m\geq 2$, then $b_{3}=b_{4}=\cdots=b_{m-1}=0$.

For the case when $r=1$ and $l=0$, if $b_{m-1}\neq 0$ then there must exist some $u$ in $S$ with $wt(u)=4m-3$ as well as some $v$ in $S$ with $wt(v)=5$. But then we have $u+v\in C$ with $wt(u+v) \leq 4m+2$, a contradiction to the minimum weight of $C$. Then we must have $b_{m-1}=0$. Then by (\ref{eq1}) we have \[ c_{2m+1}=\alpha_{2m+1,0}=\beta_{2m+1,1}+\Sigma_{j=m}^{m-1}\beta_{2m+1,j}b_{j}. \]
This gives us $c_{2m+1}=\alpha_{2m+1,0}=\beta_{2m+1,1}$. The $\alpha_{ij}$ and $\beta_{ij}$ were computed in \cite{BW12} and so we get \[ -\frac{(12m+1)(56m+4)}{(2m+1)(m-1)}\binom{5m-1}{m-2}=-2^{5}\frac{3m-1}{2m+1}\binom{5m-1}{m-2},
\] which has no integer solution.

For the case when $r=1$ and $l=1$, again we must have $b_{m-1}=0$. Then (\ref{eq1}) gives us \[\alpha_{2m+1,0}=\beta_{2m+1,1}+\beta_{2m+1,m}b_{m}, \] and so
\begin{equation}\label{eq2} b_{m}=\frac{\alpha_{2m+1,0}-\beta_{2m+1,1}}{\beta_{2m+1,m}}=\frac{-12m+5}{-4m-2}\binom{5m+1}{m}-\frac{3m}{2m+1}\binom{5m}{m-1}, \end{equation} which must be an integer for such a code to exist.

For the case when $r=2$ and $l=0$ we have $b_{2}=b_{3}=\cdots=b_{m-2}=0$ and from (\ref{eq1}) we get \[ \alpha_{2m+1,0}=\beta_{2m+1,1}+\beta_{2m+1,m-1}b_{m-1},\]  and so
\begin{equation}\label{eq3} b_{m-1}=\frac{\alpha_{2m+1,0}-\beta_{2m+1,1}}{\beta_{2m+1,m-1}}=\frac{2(6m+1)(8m+1)}{16m(2m+1)}\binom{5m}{m-1}-\frac{3m-1}{2m+1}\binom{5m-1}{m-2}, \end{equation}
which must be an integer for such a code to exist.

When $r=3$ and $l=0$, we have $b_{2}=b_{3}=\cdots=b_{m-2}=0$ and from (\ref{eq1}) we get \[\alpha_{2m+1,0}=\beta_{2m+1,1}+\beta_{2m+1,m-1}b_{m-1}, \] which gives \begin{equation}\label{eq4} b_{m-1}=\frac{\alpha_{2m+1,0}-\beta_{2m+1,1}}{\beta_{2m+1,m-1}}=\frac{3(4m+1)(6m+1)}{8m(2m+1)}\binom{5m}{m-1}-\frac{3m-1}{2m+1}\binom{5m-1}{m-2}, \end{equation} which must be an integer for such a code to exist.
\end{proof}

For the near extremal self-dual code, we have a similar result.
\begin{theorem}\label{th2.2} A near extremal self-dual code with minimal shadow does not exist whenever:
\begin{enumerate}
\item $r=1$, $l=0$ and $\frac{24m+2}{m}\binom{5m-1}{m-1}-\frac{3}{2}\binom{5m-1}{m}$ is not an integer,
\item $r=2$, $l=0$ and $\frac{24m+4}{m}\left[\binom{5m}{m-2}+3\binom{5m+1}{m-2}\right]-\frac{3}{2}\binom{5m-1}{m}$ is not an integer.
\end{enumerate}
\end{theorem}
\begin{proof}
Let $C$ be a near extremal self-dual code with minimal shadow. Then we have $d=4m+2$ and $wt(S)=r$ for $r=1,2,3$ and $wt(S)=4$ for $r=0$, and $a_{0}=1,\ a_{1}=a_{2}=\cdots=a_{2m}=0$.

 If $r>0$, then by Theorem~\ref{shadow}, we have $b_{0}=1$ for $m\geq 1$, and $b_{0}=0$ and $b_{1}=1$ for $r=0$ and $m\geq 2$. If $r>0$ and $m \geq 1$ then $b_{1}=b_{2}=\cdots=b_{m-2}=0$, otherwise there will be $v$ in $S$ with $wt(v) \leq 4m-8+r$, and $u$ in $S$ with $wt(u)=r$ so that $u+v$ is in $C$ and $wt(u+v) \leq 4m-8+2r \leq 4m-2$, a contradiction to the minimum weight of $C$. Similarly, if $r=0$ and $m \geq 2$ then $b_{2}=\cdots=b_{m-2}=0$.

Now suppose $r=0,1,2$ and $l=0$. If $b_{m-1} \neq 0$ there will be $u$ and $v$ in $S$ with $wt(u+v) \leq 4m-3+r \leq 4m$, a contradiction to the minimum weight of $C$. Then $b_{m-1}=0$. From (\ref{eq1}) we have \[\alpha_{2m,0}=\beta_{2m,\epsilon}+\beta_{2m,m}, \] where $\epsilon=0$ if $r>0$ and $\epsilon=1$ if $r=0$. According to \cite{R98} we have
 \begin{align*}
\alpha_{2m}(24m+2r)&=-\frac{12m+r}{2m}[\textup{coeff. of }y^{2m-1}\textup{ in }(1+y)^{-4m-r-1}(1-y)^{-4m}]\\
          &= -\frac{12m+r}{2m}[\textup{coeff. of }y^{2m-1}\textup{ in }(1+y)^{-r-1}(1-y^{2})^{-4m}]\\
          &=-\frac{12m+r}{2m}[\textup{coeff. of }y^{2m-1}\textup{ in }(1-y^{2})^{-4m-r-1}(1-y)^{r+1}]\\
          &=-\frac{12m+r}{2m}[\textup{coeff. of }y^{2m-1}\textup{ in }(1-y)\Sigma_{j=0}^{m}\binom{4m+r+j}{j}y^{2j}]\\
          &=\begin{cases}-\frac{12m+1}{m}\binom{5m-1}{m-1}; &\textup{ if } r=1,\\
            \frac{6m+1}{m}[\binom{5m}{m-2}+3\binom{5m+1}{m-1}]; &\textup{ if }r=2.\end{cases}
\end{align*}
 We also have $\beta_{2m,0}=2^{-r}\frac{3}{2}\binom{5m-1}{m}$ and $\beta_{2m,m}=2^{-r}$. Then if $r=1$, (\ref{eq1}) gives us \begin{equation}\label{eq5} b_{m}=\frac{24m+2}{m}\binom{5m-1}{m-1}-\frac{3}{2}\binom{5m-1}{m}, \end{equation} which must be an integer for such a code to exist, and if $r=2$, (\ref{eq1}) gives us \begin{equation}\label{eq6} b_{m}=\frac{24m+4}{m}\left[\binom{5m}{m-2}+3\binom{5m+1}{m-2}\right]-\frac{3}{2}\binom{5m-1}{m}, \end{equation} which must also be an integer for a code to exist.
\end{proof}

If $C$ is an extremal self-dual code of length $24m+8l+2r$ with near near minimal shadow we get by a similar argument as above that \[ b_{m-1}=2^{-5}\frac{(12m+1)(56m+4)}{(2m+1)(m-1)}\binom{5m-1}{m-2},\] whence the following.
\begin{theorem}\label{th2.3} An extremal self-dual code of length $24m+8l+2r$ with near near minimal shadow does not exist whenever $r=1$ and $l=0$ and $2^{-5}\frac{(12m+1)(56m+4)}{(2m+1)(m-1)}\binom{5m-1}{m-2}$ is not an integer.
\end{theorem}

We will also make use of the following lemma, which was originally proved by Ray-Cahaudhuri and Wilson in \cite{RW75}.

\begin{lemma}\label{le2.4} Let $X$ be a set of cardinality $v$. For $s \leq k \leq v-s$ let $\mathfrak{B}$ be a collection of subsets of $X$ each having cardinality $k$ and having the property that, for $B,B' \in \mathfrak{B}$, $B \neq B'$, the cardinality of $B \cap B'$ takes only $s$ distinct values. Then $|\mathfrak{B}| \leq \binom{v}{s}$.
\end{lemma}
\begin{remark}\label{re4.5}
Let $C$ be a self-dual code of length $n=24m+8l+2r$ with $m \geq 2$ not having minimal shadow, let $s:=wt(S)$ and denote the set of vectors of $S$ of minimum weight by $B_{s}$. Suppose that $2s-d \leq 2$. It follows that if $u$ and $v$ are members of $B_{s}$, then $wt(u \cap v) \leq 1$. If $2s-d = 2$ then the members of $S$ of minimum weight can intersect in either $0$ or $1$ nonzero coordinate positions. Because of the orthogonality relations among the cosets of $C_{0}$ in $C_{0}^{\perp}$, i.e. since $C_{1} \perp C_{3}$ and $C_{i} \not\perp C_{i}$ for $i=1,3$, we have any two members of $C_{i}$ intersecting in one nonzero coordinate position for $i = 1,2$. We also have that if $u \in C_{1}$ and $v \in C_{3}$ then $wt(u \cap v)=0$. Let $\mathfrak{B}_{i}$ be the set of vectors in $C_{i}$ of weight $s$. Then we have $\mathfrak{B}_{1}$ and $\mathfrak{B}_{3}$ are disjoint. Let $m_{i}$ be the effective length of $\mathfrak{B}_{i}$. Then by Lemma~\ref{le2.4} we have $B_{s} \leq m_{1}+m_{3} \leq n$.
\end{remark}

\subsection{Application to Self-Dual Codes of Lengths $74,\ 76,\ 82,\ 98,$ and $100$}

In \cite{DGH97} several weight enumerators are computed for binary singly even self-dual codes of length $n$ for $66 \leq n \leq 100$. For each length they give a combination of weight enumerators for that of a code with minimal, near minimal, and near near minimal shadow. We have eliminated several of the possibilities by using (\ref{eq2})-(\ref{eq6}) either to show the value is not an integer, or that it does not agree with the value computed in \cite{DGH97}. For $n=74$ and $n=98$ we get resp. $5447/3$ and $38301/2$ as the value $b_{m}$ and so Part 1 of Theorem \ref{th2.2} applies. For $n=76,\ 82,\ 100$ we use resp. (\ref{eq6}), (\ref{eq2}), (\ref{eq6}) to get values (Table \ref{bm3}) that do not agree with those given in \cite{DGH97}, which were computed using the method introduced by Conway and Sloane in \cite{CS90}. We also use the comment following Lemma~\ref{le2.4} to narrow the possible range for the parameter in the near near extremal weight enumerators for cases $n=82$ and $100$. These restrictions are summarized in Tables \ref{bm1} and \ref{bm2} below.

\begin{table}[ht]
\caption{Summary of restrictions on possible weight enumerator for lengths $74,76,82,98,100$}
\centering
\begin{tabular}{|c|c|c|c|}
\hline\hline
$n$ & Weight Enumerator Eliminated & $b_{m}$ & Reference\\[0.5ex]
\hline \hline
74&Minimal Shadow& $5447/3$ &Part 1 of Theorem \ref{th2.2}\\ \hline
76&Minimal Shadow& $1050$ & Equation (\ref{eq6})\\ \hline
82&Near Minimal Shadow& $1105$ & Equation (\ref{eq2})\\ \hline
98&Minimal Shadow&$38301/2$&Part 1 of Theorem \ref{th2.2}\\ \hline
100&Minimal Shadow&14686& Equation (\ref{eq2})\\ [1ex]
\hline
\end{tabular}
\label{bm1}
\end{table}

\begin{table}[ht]
\caption{Summary of restrictions on possible range for $\alpha,\beta$ in the near near minimal shadow case for lengths $82,100$ }
\centering
\begin{tabular}{|c|c|c|c|}
\hline\hline
$n$ & New range for $\alpha,\beta$ & Reference\\[0.5ex]
\hline \hline
82& $0\leq \alpha \leq 82$ & Remark \ref{re4.5}\\ \hline

100&$0\leq\alpha\leq min\{100,-\frac{1}{20}\beta\}\textup{ where }-3265\leq \beta\leq 0$&Remark \ref{re4.5}\\ [1ex]
\hline
\end{tabular}
\label{bm2}
\end{table}

\begin{table}[ht]
\caption{Contradictory values of $b_{m}$ for cases $n=76,82$ and $100$}
\centering
\begin{tabular}{|c|c|c|}
\hline\hline
$n$ & $b_{m}$ computed using above method & $b_{m}$ computed using method of \cite{CS90} \\[0.5ex]
\hline
76&1050&2590\\ \hline
82&1105&1505\\ \hline
100&14686&98686\\ [1ex]
\hline
\end{tabular}
\label{bm3}
\end{table}

We now list the possible weight enumerators of extremal and near extremal singly even self-dual codes of lengths $n=74,\ 76,\ 82,\ 98,$ and $100$.
\begin{itemize}
  \item The possible weight enumerators for self-dual $[74,37,14]$ codes are
   \[\begin{cases}S_{1}=-\alpha y^{9}+(2590+14\alpha)y^{13}+(674584-91\alpha)y^{17}+(364\alpha+44035772)y^{21}+\cdots,\\
     W_{1}=1+(6364+32\alpha)y^{14}+(100603-160\alpha)y^{16}+(32\alpha+1061678)y^{18}+\cdots,\\
     (-185\leq\alpha\leq 0), \end{cases}\]
  and
  \[\begin{cases}S_{2}=y^{5}+(-16-\alpha)y^{9}+(2710+14\alpha)y^{13}+(674024-91\alpha)y^{17}+\cdots,\\
 W_{2}=1+(6346+320\alpha)y^{14}+(102651-160\alpha)y^{16}+(32\alpha+1039150)y^{18}+\cdots,\\
 (-19\leq\alpha\leq-16).\end{cases}\]
 The weight enumerator for the minimal shadow case was eliminated in this paper. There is no known code for either case.
  \item The possible weight enumerators for self-dual $[76,38,14]$ codes are
  \[\begin{cases}S_{1}=\alpha y^{10}+(9500-14\alpha)y^{14}+(1831600+91\alpha)y^{18}+(105689400-364\alpha)y^{22}+\cdots,\\
  W_{1}=1+(4750-16\alpha)y^{14}+(79895+64\alpha)y^{16}+(64\alpha+915800)y^{18}+\cdots,\\
  (0\leq\alpha\leq 296),\end{cases}\]
 and
  \[\begin{cases}S_{2}=y^{6}+(-16-\alpha)y^{10}+(9620+14\alpha)y^{14}+(1831040-91\alpha)y^{18}+\cdots,\\
  W_{2}=1+(4750+16\alpha)y^{14}+(80919-64\alpha)y^{16}+(905560-64\alpha)y^{18}+\cdots,\\
  (-296\leq\alpha\leq-16).\end{cases}\]
  The weight enumerator for the minimal shadow case was eliminated in this paper. In \cite{BY03}, a code with weight enumerator $W_{1}$ for $\alpha=0$ was constructed by assuming an automorphism of order $19$. It is shown in \cite{DY03} that there are exactly three inequivalent self-dual $[76,38,14]$ codes having an automorphism of order $19$. All of these have weight enumerator $W_{1}$ with $\alpha=0$.
  \item The possible weight enumerator for self-dual $[82,41,16]$ codes is
  \[\begin{cases}S_{1}=\alpha y^{9}+(1640-\alpha)y^{13}+(281424+120\alpha)y^{17}+(-560\alpha+33442552)y^{21}+\cdots,\\
  W_{1}=1+(39524+128\alpha)y^{16}+(556985-896\alpha)y^{18}+(1536\alpha+5628480)y^{20}+\cdots,\\
  (0\leq\alpha\leq 82).\end{cases}\]

  The weight enumerator for the near minimal shadow case was eliminated, and the range for the parameter in the near near minimal shadow case was improved in this paper. There is no known code with this weight enumerator.
  \item The possible weight enumerators for self-dual $[98,49,18]$ codes are
  \[\begin{cases}S_{1}=\alpha y^{9}+(-\beta-20\alpha)y^{13}+(190\alpha+18\beta+27930)y^{17}+(-1140\alpha-153\beta+9118816)y^{21}+\cdots,\\
  W_{1}=1+(70756+32\beta)y^{18}+(2048\alpha+1256752-160\beta)y^{20}+(-96\beta-22528\alpha+15857968)y^{22}+\cdots,\\
  (0\leq\alpha\leq min\{2,\frac{1}{20}\beta\}\textup{ where }0\leq \beta \leq 2211),\end{cases}\]
 and
  \[\begin{cases}S_{2}=y^{5}+(-209-\alpha)y^{13}+(30570+18\alpha)y^{17}+(9101051-153\alpha)y^{21}+\cdots,\\
  W_{2}=1+(70756+32\alpha)y^{18}+(1301808-16\alpha)y^{20}+(-96\alpha+15231280)y^{22}+\cdots,\\
  (-1698\leq\alpha\leq-209).\end{cases}\]
  The weight enumerator for the minimal shadow case was eliminated in this paper. The range for the parameter in the near near minimal shadow case was improved in \cite{HM06}. There is no known code for either case.
  \item The possible weight enumerators for self-dual $[100,50,18]$ codes are
  \[\begin{cases}S_{1}=\alpha y^{10}+(-\beta-20\alpha)y^{14}+(18\beta+104500-190\alpha)y^{18}+(-153\beta-1140\alpha+26155200)y^{22}+\cdots,\\
  W_{1}=1+(16\beta+52250)y^{18}+(972180-64\beta+1024\alpha)y^{20}+(13077600-128\beta-10240\alpha)y^{22}+\cdots,\\
  (0\leq\alpha\leq min\{100,-\frac{1}{20}\beta\}\textup{ where }-3265\leq \beta\leq 0),\end{cases}\]
and
  \[\begin{cases}S_{2}=y^{6}+(-209-\alpha)y^{14}+(107140+18\alpha)y^{18}+(26137435-153\alpha)y^{22}+\cdots,\\
  W_{2}=1+(52250+16\alpha)y^{18}+(994708-64\alpha)y^{20}+(-128\alpha+12786784)y^{22}+\cdots,\\
  (-5952\leq\alpha\leq-209).\end{cases}\]
   The weight enumerator for the near minimal shadow case was eliminated, and the range for the parameter in the near near minimal shadow case was improved in this paper. There is no known code for either case.
\end{itemize}

\section{Conclusion}\label{cl}
This paper demonstrates some results on self-dual codes. We make two contributions to this topic. The first one is the decomposition of binary self-dual $[4p+f,2p+\frac{f}{2},d]$ $(f=0,2,4)$ codes with dihedral automorphism group $D_{2p}$, where $p$ is an odd prime. These results are applied to classify self-dual $[78,39,14]$ codes with dihedral automorphism group $D_{38}$ and we obtain some self-dual codes with new weight enumerators. Furthermore, we also show that there are at least $141$ inequivalent self-dual $[116,58,18]$ codes with dihedral automorphism group $D_{58}$. Up to equivalence, most of these codes are new since the orders of the automorphism groups of all but one known self-dual $[116,58,18]$ code are divided by $23$. The second one is the restriction of the extremal self-dual codes with near minimal shadow, and near extremal self-dual codes with minimal, near minimal, and near near minimal shadow. And using these results, we eliminate some of the possible weight enumerators of self-dual codes with lengths $74,\ 76,\ 82,\ 98$ and $100$ determined in \cite{CS90} and \cite{DGH97}. Self-dual codes with these weight enumerators have been constructed only for the length $76$ \cite{DY03}, \cite{BY03}. Constructing the self-dual codes with these weight enumerators of other lengths seems to be a challenging problem.

\begin{center}
\begin{longtable}{|c|c|c|c|c|c|c|c|c|c|}
\caption{Self-dual $[116,58,18]$ codes with dihedral automorphism group $D_{58}$}\\\hline
\label{116}
Code  &  $u_{1}$  &  $u_{2}$  & $u_{3}$  & $v_{1}$ & $v_{2}$ & $s$ & $A_{18}$ & $I_{36}$ & $\sharp$Aut\\ \hline
\endfirsthead
\multicolumn{4}{c}%
{\tablename\ \thetable\ \textit{Continued}} \\
\hline
Code  &  $u_{1}$  &  $u_{2}$  & $u_{3}$  & $v_{1}$ & $v_{2}$ & $s$ & $A_{18}$ & $I_{36}$ & $\sharp$Aut\\ \hline
\hline
\endhead
\hline
\endlastfoot
$C_{1}$ & $9$ & $153$ & $144$ & $882$ & $12183$ & $(2,3,4)$ & $2146$ & $178205$ & $58$\\ \hline
$C_{2}$ & $37$ & $8$ & $29$ & $882$ & $12183$ & $I$ & $2378$ & $209989$  & $58$\\ \hline
$C_{3}$ & $14$ & $259$ & $273$ & $259$ & $15951$ & $(2,3,4)$ & $2610$ & $260391$  & $58$\\ \hline
$C_{4}$ & $21$ & $34$ & $55$ & $259$ & $15951$ & $(2,3,4)$ & $2784$ & $287912$  & $58$\\ \hline
$C_{5}$ & $3$ & $200$ & $203$ & $259$ & $15951$ & $(1,2,3,4)$ & $2842$ & $301397$  & $58$\\ \hline
$C_{6}$ & $116$ & $85$ & $31$ & $259$ & $15951$ & $(1,2,3,4)$ & $2842$ & $307081$  & $58$\\ \hline
$C_{7}$ & $13$ & $189$ & $176$ & $882$ & $12183$ & $(2,3,4)$ & $2842$ & $300556$  & $58$\\ \hline
$C_{8}$ & $14$ & $132$ & $118$ & $882$ & $12183$ & $I$ & $2842$ & $305196$  & $58$\\ \hline
$C_{9}$ & $28$ & $134$ & $106$ & $259$ & $15951$ & $(2,3,4)$ & $2842$ & $299396$  & $58$\\ \hline
$C_{10}$ & $2$ & $138$ & $140$ & $882$ & $12183$ & $(1,2,3,4)$ & $2900$ & $313287$  & $58$\\ \hline
$C_{11}$ & $19$ & $99$ & $118$ & $882$ & $12183$ & $(1,2,3,4)$ & $2900$ & $318565$  & $58$\\ \hline
$C_{12}$ & $19$ & $99$ & $118$ & $882$ & $12183$ & $(2,3,4)$ & $2900$ & $310880$  & $58$\\ \hline
$C_{13}$ & $13$ & $145$ & $158$ & $259$ & $15951$ & $(1,2,3,4)$ & $2900$ & $306066$  & $58$\\ \hline
$C_{14}$ & $37$ & $33$ & $4$ & $882$ & $12183$ & $I$ & $2900$ & $312417$  & $58$\\ \hline
$C_{15}$ & $1$ & $156$ & $155$ & $882$ & $12183$ & $(2,3,4)$ & $2900$ & $315549$  & $58$\\ \hline
$C_{16}$ & $29$ & $143$ & $172$ & $882$ & $12183$ & $(1,2,3,4)$ & $2958$ & $325119$  & $58$\\ \hline
$C_{17}$ & $23$ & $169$ & $146$ & $882$ & $12183$ & $(2,3,4)$ & $2958$ & $327410$  & $58$\\ \hline
$C_{18}$ & $17$ & $39$ & $56$ & $882$ & $12183$ & $(2,3,4)$ & $3016$ & $343360$  & $58$\\ \hline
$C_{19}$ & $272$ & $245$ & $27$ & $5469$ & $9024$ & $(1,2,3,4)$ & $3016$ & $342171$  & $58$\\ \hline
$C_{20}$ & $44$ & $39$ & $5$ & $259$ & $15951$ & $I$ & $3016$ & $340547$  & $58$\\ \hline
$C_{21}$ & $5$ & $234$ & $229$ & $882$ & $12183$ & $(2,3,4)$ & $3016$ & $341620$  & $58$\\ \hline
$C_{22}$ & $21$ & $120$ & $99$ & $882$ & $12183$ & $(2,3,4)$ & $3016$ & $337995$  & $58$\\ \hline
$C_{23}$ & $5$ & $150$ & $155$ & $5469$ & $9024$ & $(1,2,3,4)$ & $3074$ & $342983$  & $58$\\ \hline
$C_{24}$ & $29$ & $200$ & $229$ & $882$ & $12183$ & $(2,3,4)$ & $3074$ & $358933$  & $58$\\ \hline
$C_{25}$ & $14$ & $96$ & $110$ & $259$ & $15951$ & $(2,3,4)$ & $3074$ & $348174$  & $58$\\ \hline
$C_{26}$ & $97$ & $67$ & $30$ & $882$ & $12183$ & $I$ & $3074$ & $356903$  & $58$\\ \hline
$C_{27}$ & $10$ & $167$ & $157$ & $5469$ & $9024$ & $I$ & $3074$ & $348377$  & $58$\\ \hline
$C_{28}$ & $5$ & $279$ & $284$ & $882$ & $12183$ & $(2,3,4)$ & $3132$ & $361717$  & $58$\\ \hline
$C_{29}$ & $10$ & $83$ & $93$ & $5469$ & $9024$ & $(2,3,4)$ & $3132$ & $372186$  & $58$\\ \hline
$C_{30}$ & $39$ & $317$ & $278$ & $882$ & $12183$ & $(2,3,4)$ & $3132$ & $368793$  & $58$\\ \hline
$C_{31}$ & $31$ & $25$ & $6$ & $882$ & $12183$ & $I$ & $3132$ & $359716$  & $58$\\ \hline
$C_{32}$ & $16$ & $20$ & $4$ & $882$ & $12183$ & $(2,3,4)$ & $3132$ & $367169$  & $58$\\ \hline
$C_{33}$ & $2$ & $265$ & $267$ & $259$ & $15951$ & $(2,3,4)$ & $3190$ & $381495$  & $58$\\ \hline
$C_{34}$ & $27$ & $83$ & $110$ & $259$ & $15951$ & $(1,2,3,4)$ & $3190$ & $371809$  & $58$\\ \hline
$C_{35}$ & $35$ & $14$ & $49$ & $259$ & $15951$ & $(1,2,3,4)$ & $3190$ & $374593$  & $58$\\ \hline
$C_{36}$ & $36$ & $134$ & $170$ & $259$ & $15951$ & $(2,3,4)$ & $3190$ & $381031$  & $58$\\ \hline
$C_{37}$ & $198$ & $185$ & $13$ & $5469$ & $9024$ & $I$ & $3190$ & $382916$  & $58$\\ \hline
$C_{38}$ & $44$ & $29$ & $15$ & $882$ & $12183$ & $I$ & $3190$ & $373375$  & $58$\\ \hline
$C_{39}$ & $136$ & $105$ & $31$ & $259$ & $15951$ & $(1,2,3,4)$ & $3190$ & $382568$  & $58$\\ \hline
$C_{40}$ & $9$ & $189$ & $180$ & $5469$ & $9024$ & $I$ & $3190$ & $378276$  & $58$\\ \hline
$C_{41}$ & $10$ & $167$ & $157$ & $5469$ & $9024$ & $(2,3,4)$ & $3190$ & $382104$  & $58$\\ \hline
$C_{42}$ & $12$ & $259$ & $247$ & $5469$ & $9024$ & $(2,3,4)$ & $3190$ & $391123$  & $58$\\ \hline
$C_{43}$ & $22$ & $166$ & $188$ & $5469$ & $9024$ & $(2,3,4)$ & $3248$ & $388455$  & $58$\\ \hline
$C_{44}$ & $42$ & $16$ & $58$ & $882$ & $12183$ & $(2,3,4)$ & $3248$ & $386280$  & $58$\\ \hline
$C_{45}$ & $3$ & $200$ & $203$ & $259$ & $15951$ & $(2,3,4)$ & $3248$ & $396778$  & $58$\\ \hline
$C_{46}$ & $201$ & $180$ & $21$ & $259$ & $15951$ & $(1,2,3,4)$ & $3248$ & $389847$  & $58$\\ \hline
$C_{47}$ & $12$ & $259$ & $247$ & $5469$ & $9024$ & $I$ & $3248$ & $391645$  & $58$\\ \hline
$C_{48}$ & $13$ & $189$ & $176$ & $882$ & $12183$ & $I$ & $3248$ & $392022$  & $58$\\ \hline
$C_{49}$ & $4$ & $172$ & $176$ & $5469$ & $9024$ & $(1,2,3,4)$ & $3306$ & $406522$  & $58$\\ \hline
$C_{50}$ & $40$ & $217$ & $257$ & $5469$ & $9024$ & $(1,2,3,4)$ & $3306$ & $408958$  & $58$\\ \hline
$C_{51}$ & $40$ & $217$ & $257$ & $5469$ & $9024$ & $(2,3,4)$ & $3306$ & $408697$  & $58$\\ \hline
$C_{52}$ & $44$ & $29$ & $15$ & $882$ & $12183$ & $(1,2,3,4)$ & $3306$ & $398750$  & $58$\\ \hline
$C_{53}$ & $9$ & $153$ & $144$ & $882$ & $12183$ & $I$ & $3306$ & $412119$  & $58$\\ \hline
$C_{54}$ & $21$ & $120$ & $99$ & $882$ & $12183$ & $I$ & $3306$ & $412554$  & $58$\\ \hline
$C_{55}$ & $23$ & $169$ & $146$ & $882$ & $12183$ & $I$ & $3306$ & $404434$  & $58$\\ \hline
$C_{56}$ & $5$ & $279$ & $284$ & $882$ & $12183$ & $I$ & $3335$ & $412815$  & $58$\\ \hline
$C_{57}$ & $10$ & $83$ & $93$ & $5469$ & $9024$ & $(1,2,3,4)$ & $3364$ & $417890$  & $58$\\ \hline
$C_{58}$ & $22$ & $166$ & $188$ & $5469$ & $9024$ & $(1,2,3,4)$ & $3364$ & $413830$  & $58$\\ \hline
$C_{59}$ & $29$ & $200$ & $229$ & $882$ & $12183$ & $(1,2,3,4)$ & $3364$ & $417165$  & $58$\\ \hline
$C_{60}$ & $208$ & $198$ & $10$ & $5469$ & $9024$ & $I$ & $3364$ & $428098$  & $58$\\ \hline
$C_{61}$ & $272$ & $245$ & $27$ & $5469$ & $9024$ & $I$ & $3364$ & $425778$  & $58$\\ \hline
$C_{62}$ & $5$ & $234$ & $229$ & $882$ & $12183$ & $I$ & $3364$ & $423777$  & $58$\\ \hline
$C_{63}$ & $2$ & $138$ & $140$ & $882$ & $12183$ & $(2,3,4)$ & $3422$ & $435812$  & $58$\\ \hline
$C_{64}$ & $6$ & $172$ & $278$ & $882$ & $12183$ & $(1,2,3,4)$ & $3422$ & $428939$  & $58$\\ \hline
$C_{65}$ & $39$ & $272$ & $311$ & $5469$ & $9024$ & $(2,3,4)$ & $3422$ & $442511$  & $58$\\ \hline
$C_{66}$ & $42$ & $55$ & $97$ & $882$ & $12183$ & $(1,2,3,4)$ & $3422$ & $438045$  & $58$\\ \hline
$C_{67}$ & $125$ & $122$ & $3$ & $5469$ & $9024$ & $I$ & $3422$ & $442395$  & $58$\\ \hline
$C_{68}$ & $17$ & $49$ & $66$ & $5469$ & $9024$ & $(2,3,4)$ & $3480$ & $449007$  & $58$\\ \hline
$C_{69}$ & $42$ & $55$ & $97$ & $882$ & $12183$ & $(2,3,4)$ & $3480$ & $452284$  & $58$\\ \hline
$C_{70}$ & $2$ & $265$ & $267$ & $259$ & $15951$ & $(1,2,3,4)$ & $3480$ & $445556$  & $58$\\ \hline
$C_{71}$ & $27$ & $83$ & $110$ & $259$ & $15951$ & $(2,3,4)$ & $3480$ & $458200$  & $58$\\ \hline
$C_{72}$ & $184$ & $165$ & $19$ & $5469$ & $9024$ & $I$ & $3480$ & $454778$  & $58$\\ \hline
$C_{73}$ & $140$ & $118$ & $22$ & $5469$ & $9024$ & $I$ & $3480$ & $447992$  & $58$\\ \hline
$C_{74}$ & $37$ & $8$ & $29$ & $882$ & $12183$ & $(1,2,3,4)$ & $3480$ & $447325$  & $58$\\ \hline
$C_{75}$ & $5$ & $150$ & $155$ & $5469$ & $9024$ & $(2,3,4)$ & $3538$ & $470641$  & $58$\\ \hline
$C_{76}$ & $34$ & $227$ & $263$ & $882$ & $12183$ & $(2,3,4)$ & $3538$ & $464638$  & $58$\\ \hline
$C_{77}$ & $44$ & $237$ & $281$ & $882$ & $12183$ & $(1,2,3,4)$ & $3538$ & $464928$  & $58$\\ \hline
$C_{78}$ & $14$ & $132$ & $118$ & $882$ & $12183$ & $(2,3,4)$ & $3538$ & $463594$  & $58$\\ \hline
$C_{79}$ & $210$ & $190$ & $20$ & $882$ & $12183$ & $I$ & $3596$ & $484010$  & $58$\\ \hline
$C_{80}$ & $91$ & $63$ & $28$ & $259$ & $15951$ & $(1,2,3,4)$ & $3596$ & $475455$  & $58$\\ \hline
$C_{81}$ & $9$ & $189$ & $180$ & $5469$ & $9024$ & $(2,3,4)$ & $3596$ & $486214$  & $58$\\ \hline
$C_{82}$ & $1$ & $156$ & $155$ & $882$ & $12183$ & $I$ & $3596$ & $478645$  & $58$\\ \hline
$C_{83}$ & $39$ & $272$ & $311$ & $5469$ & $9024$ & $(1,2,3,4)$ & $3654$ & $489346$  & $58$\\ \hline
$C_{84}$ & $34$ & $227$ & $263$ & $882$ & $12183$ & $(1,2,3,4)$ & $3654$ & $495581$  & $58$\\ \hline
$C_{85}$ & $44$ & $237$ & $281$ & $882$ & $12183$ & $(2,3,4)$ & $3654$ & $494943$  & $58$\\ \hline
$C_{86}$ & $14$ & $259$ & $273$ & $259$ & $15951$ & $(1,2,3,4)$ & $3654$ & $495900$  & $58$\\ \hline
$C_{87}$ & $125$ & $122$ & $3$ & $5469$ & $9024$ & $(1,2,3,4)$ & $3654$ & $509820$  & $58$\\ \hline
$C_{88}$ & $184$ & $165$ & $19$ & $5469$ & $9024$ & $(1,2,3,4)$ & $3654$ & $497089$  & $58$\\ \hline
$C_{89}$ & $210$ & $190$ & $20$ & $882$ & $12183$ & $(2,3,4)$ & $3683$ & $516171$  & $58$\\ \hline
$C_{90}$ & $21$ & $34$ & $55$ & $259$ & $15951$ & $I$ & $3712$ & $499264$  & $58$\\ \hline
$C_{91}$ & $35$ & $14$ & $49$ & $259$ & $15951$ & $(2,3,4)$ & $3712$ & $509095$  & $58$\\ \hline
$C_{92}$ & $140$ & $118$ & $22$ & $5469$ & $9024$ & $(1,2,3,4)$ & $3712$ & $509588$  & $58$\\ \hline
$C_{93}$ & $31$ & $25$ & $6$ & $882$ & $12183$ & $(1,2,3,4)$ & $3712$ & $519970$  & $58$\\ \hline
$C_{94}$ & $201$ & $180$ & $21$ & $259$ & $15951$ & $I$ & $3712$ & $516084$  & $58$\\ \hline
$C_{95}$ & $136$ & $105$ & $31$ & $259$ & $15951$ & $I$ & $3712$ & $519390$ & $58$\\ \hline
$C_{96}$ & $39$ & $278$ & $317$ & $882$ & $12183$ & $(1,2,3,4)$ & $3770$ & $526727$ & $58$\\ \hline
$C_{97}$ & $42$ & $16$ & $58$ & $882$ & $12183$ & $(1,2,3,4)$ & $3770$ & $528757$ & $58$\\ \hline
$C_{98}$ & $208$ & $198$ & $10$ & $5469$ & $9024$ & $(2,3,4)$ & $3770$ & $527017$ & $58$\\ \hline
$C_{99}$ & $15$ & $2$ & $13$ & $882$ & $12183$ & $I$ & $3770$ & $536500$ & $58$\\ \hline
$C_{100}$ & $4$ & $172$ & $176$ & $5469$ & $9024$ & $(2,3,4)$ & $3828$ & $546853$ & $58$\\ \hline
$C_{101}$ & $116$ & $85$ & $31$ & $259$ & $15951$ & $I$ & $3828$ & $539255$ & $58$\\ \hline
$C_{102}$ & $6$ & $272$ & $278$ & $882$ & $12183$ & $I$ & $3915$ & $565529$ & $58$\\ \hline
$C_{103}$ & $14$ & $96$ & $110$ & $259$ & $15951$ & $I$ & $3944$ & $572228$ & $58$\\ \hline
$C_{104}$ & $208$ & $198$ & $10$ & $5469$ & $9024$ & $(1,2,3,4)$ & $3944$ & $581392$ & $58$\\ \hline
$C_{105}$ & $17$ & $49$ & $66$ & $5469$ & $9024$ & $(1,2,3,4)$ & $4002$ & $588816$ & $58$\\ \hline
$C_{106}$ & $184$ & $165$ & $19$ & $5469$ & $9024$ & $(2,3,4)$ & $4002$ & $598444$ & $58$\\ \hline
$C_{107}$ & $16$ & $20$ & $4$ & $882$ & $12183$ & $I$ & $4002$ & $605346$ & $58$\\ \hline
$C_{108}$ & $6$ & $272$ & $278$ & $882$ & $12183$ & $(2,3,4)$ & $4060$ & $605201$ & $58$\\ \hline
$C_{109}$ & $17$ & $39$ & $56$ & $882$ & $12183$ & $(1,2,3,4)$ & $4060$ & $616279$ & $58$\\ \hline
$C_{110}$ & $14$ & $96$ & $110$ & $259$ & $15951$ & $(1,2,3,4)$ & $4060$ & $616047$ & $58$\\ \hline
$C_{111}$ & $15$ & $2$ & $13$ & $882$ & $12183$ & $(1,2,3,4)$ & $4060$ & $606941$ & $58$\\ \hline
$C_{112}$ & $36$ & $134$ & $170$ & $259$ & $15951$ & $(1,2,3,4)$ & $4118$ & $635274$ & $58$\\ \hline
$C_{113}$ & $125$ & $122$ & $3$ & $5469$ & $9024$ & $(2,3,4)$ & $4147$ & $632026$ & $58$\\ \hline
$C_{114}$ & $39$ & $278$ & $317$ & $882$ & $12183$ & $I$ & $4176$ & $645511$ & $58$\\ \hline
$C_{115}$ & $299$ & $273$ & $26$ & $5469$ & $9024$ & $I$ & $4176$ & $636724$ & $58$\\ \hline
$C_{116}$ & $37$ & $33$ & $4$ & $882$ & $12183$ & $(1,2,3,4)$ & $4176$ & $647309$ & $58$\\ \hline
$C_{117}$ & $34$ & $227$ & $263$ & $882$ & $12183$ & $I$ & $4205$ & $658155$ & $58$\\ \hline
$C_{118}$ & $13$ & $145$ & $158$ & $259$ & $15951$ & $(2,3,4)$ & $4234$ & $686082$ & $58$\\ \hline
$C_{119}$ & $97$ & $67$ & $30$ & $882$ & $12183$ & $(1,2,3,4)$ & $4234$ & $661722$ & $58$\\ \hline
$C_{120}$ & $44$ & $39$ & $5$ & $259$ & $15951$ & $(1,2,3,4)$ & $4234$ & $672278$ & $58$\\ \hline
$C_{121}$ & $4$ & $172$ & $176$ & $5469$ & $9024$ & $I$ & $4292$ & $684951$ & $58$\\ \hline
$C_{122}$ & $42$ & $16$ & $58$ & $882$ & $12183$ & $I$ & $4292$ & $678803$ & $58$\\ \hline
$C_{123}$ & $15$ & $2$ & $13$ & $882$ & $12183$ & $(2,3,4)$ & $4292$ & $691592$ & $58$\\ \hline
$C_{124}$ & $210$ & $190$ & $20$ & $882$ & $12183$ & $(1,2,3,4)$ & $4292$ & $677585$ & $58$\\ \hline
$C_{125}$ & $35$ & $14$ & $49$ & $259$ & $15951$ & $I$ & $4321$ & $692114$ & $58$\\ \hline
$C_{126}$ & $29$ & $143$ & $172$ & $882$ & $12183$ & $(2,3,4)$ & $4350$ & $715169$ & $58$\\ \hline
$C_{127}$ & $198$ & $185$ & $13$ & $5469$ & $9024$ & $(1,2,3,4)$ & $4408$ & $730046$ & $58$\\ \hline
$C_{128}$ & $13$ & $145$ & $158$ & $259$ & $15951$ & $I$ & $4437$ & $725377$ & $58$\\ \hline
$C_{129}$ & $198$ & $185$ & $13$ & $5469$ & $9024$ & $(2,3,4)$ & $4437$ & $736078$ & $58$\\ \hline
$C_{130}$ & $31$ & $25$ & $6$ & $882$ & $12183$ & $(2,3,4)$ & $4553$ & $764933$ & $58$\\ \hline
$C_{131}$ & $12$ & $259$ & $247$ & $5469$ & $9024$ & $(1,2,3,4)$ & $4553$ & $784682$ & $58$\\ \hline
$C_{132}$ & $28$ & $134$ & $106$ & $259$ & $15951$ & $I$ & $4582$ & $778360$ & $58$\\ \hline
$C_{133}$ & $5$ & $279$ & $284$ & $882$ & $12183$ & $(1,2,3,4)$ & $4698$ & $817713$ & $58$\\ \hline
$C_{134}$ & $29$ & $200$ & $229$ & $882$ & $12183$ & $I$ & $4698$ & $827718$ & $58$\\ \hline
$C_{135}$ & $116$ & $85$ & $31$ & $259$ & $15951$ & $(2,3,4)$ & $4698$ & $818554$ & $58$\\ \hline
$C_{136}$ & $36$ & $134$ & $170$ & $259$ & $15951$ & $I$ & $4756$ & $838100$ & $58$\\ \hline
$C_{137}$ & $299$ & $273$ & $26$ & $5469$ & $9024$ & $(2,3,4)$ & $4785$ & $857124$ & $58$\\ \hline
$C_{138}$ & $21$ & $34$ & $55$ & $259$ & $15951$ & $(1,2,3,4)$ & $4872$ & $869536$ & $58$\\ \hline
$C_{139}$ & $37$ & $33$ & $4$ & $882$ & $12183$ & $(2,3,4)$ & $4872$ & $879715$ & $58$\\ \hline
$C_{140}$ & $3$ & $200$ & $203$ & $259$ & $15951$ & $I$ & $5075$ & $947343$ & $58$\\ \hline
$C_{141}$ & $27$ & $83$ & $110$ & $259$ & $15951$ & $I$ & $5220$ & $1005807$ & $58$\\ \hline
\end{longtable}
\end{center}

\end{document}